\documentclass{article}
\usepackage{amssymb}
\begin{document}
\title{Surface
spin-transfer torque and spin-injection effective field in
ferromagnetic junctions: \\ Unified theory}
\author{R.J. Elliott$^1$, E.M. Epshtein$^2$, Yu.V. Gulyaev$^2$, P.E.
Zilberman$^2$\thanks{e-mail:zil@ms.ire.rssi.ru} \\ \\ $^1$\textit{University of Oxford, Oxford, QX1, United
Kingdom}, \\ $^2$\textit{Institute of Radio
Engineering and Electronics, Fryazino, 141190, Russia}}
\date{}
\maketitle
\begin{abstract}
 We consider theoretically a current flowing perpendicular to
 interfaces of a spin-valve type ferromagnetic metallic junction.
 For the first time an effective approach is investigated to
 calculate a simultaneous action of the two current effects,
 namely, the nonequilibrium longitudinal spin injection and the
 transversal spin-transfer surface torque. Dispersion relation for
 fluctuations is derived and solved. Nonlinear problem is solved
 about steady state arising due to instability for a thick enough
 free layer.
\end{abstract}

\section{Introduction}\label{sec1}

Great attention is attracted now to features of current flowing
through ferromagnetic junctions of a spin-valve type, i.e.
structures with contacting ferromagnetic thin layers, one of them
having pinned and the other free spins. As experiments showed,
current can influence substantially the free layer magnetic state
of such junctions that leads to resistance
jumps~\cite{Katine}--~\cite{Wegrowe}, as well as microwave
emission~\cite{Tsoi}--~\cite{Bass}.

Mechanisms of the current effect
are not completely clear so far. A mechanism was proposed
in~\cite{Heide} of the current effect on the free ferromagnetic
layer magnetization $\mathbf{M}$ due to injection of nonequilibrium
longitudinal (i.e., collinear to $\mathbf{M}$) spins into the
layer. Detailed theory of the mechanism was developed
in~\cite{Gulyaev1}--~\cite{Gulyaev3}. The injection creates a
nonequilibrium carrier spin polarization in the layer. This
polarization, in its turn, contributes to the \textit{sd} exchange
energy $U_{sd}(\mathbf{j})$ and to the corresponding \textit{sd}
exchange effective field $\mathbf{H}_{sd}(\mathbf{j})$, which is
dependent on the electric current density $\mathbf{j}$. For current
densities exceeding some threshold, a reorientation first-order phase
transition occurs and the magnetization vector $\mathbf{M}$
direction may change abruptly. Such a current induced magnetization
reversal (or current induced switching) leads to resistance jumps,
which corresponds to the experimental
data ~\cite{Katine}--~\cite{Wegrowe}.

On the other hand, another
mechanism of current effect on the ferromagnetic layer state was
proposed~\cite{Slonczewski,Berger} long before
appearing of the experimental
results~\cite{Katine}--\cite{Bass}. According
to the mechanism, the transverse (with respect to $\mathbf{M}$)
mobile electrons spin current is to vanish near the interface
between two (pinned and free) contacting non-collinear
ferromagnets. It occurs due to statistical velocity spread of the
mobile electrons not because of the relaxation. The electrons
interact with magnetic lattice (\textit{sd} exchange interaction) and that
is why a torque appears at the interface. This spin torque acts on
the lattice and transfers the lost spin current to it. This
spin-transfer torque may exceed the dissipation torque at large
enough current densities. Then the initial state becomes unstable
and the magnetization reversal occurs. As estimations show, such a
switching mechanism can agree with experimental results~\cite{Katine}--\cite{Bass} also.

In this connection, a very significant question arises: how are
things going in the real experiments when both effects coexist,
namely, longitudinal spin injection and corresponding current
dependent effective field, and current dependent spin-transfer
torque at the interface of the magnetic junction. Up to now, these
effects were studied separately. Meanwhile, these effects do not
only coexist, but influence each other also. Therefore, both
effects are to be taken into account simultaneously, in scope of a
unified theory, to understand better the experimental situation.
Such a theory was developed for the
first time in our previous preprint \cite{Gulyaev4} and article
\cite{Gulyaev5}. The aim of the present paper is to elaborate a new effective
approach for solving the problem. This approach allows not only get
all the previously achievable results (see \cite{Gulyaev4,Gulyaev5}), but
additionally to solve nonlinear problem about the steady state
arising due to instability development. The approach based on the
interesting feature of the effective field $\mathbf{H}_{sd}$  that
was revealed in \cite{Gulyaev4,Gulyaev5}. This field is created due
to the injection of
nonequilibrium longitudinal mobile spins into the free layer volume
and depends on the bulk parameters of the free layer. But a point
of the field localization occurs near the interface. This feature
suggests a possibility to include an effect of the field into
boundary condition and shows the transversal spin flux becomes, in
fact, discontinues at the interface. The simplification arises
because of no singularity remains in the equations of motion.

\section{Model}\label{sec2}

We will consider a spin-valve type magnetic junction with current
flowing across the layer interfaces (CPP geometry). Ferromagnetic
metal layer \textbf{1} has pinned orientations of the lattice and mobile
electron spins. Another ferromagnetic metal layer \textbf{2}
contacts with the layer \textbf{1} at a point $x=0$ and has free
lattice\footnote{We mean a lattice of magnetic ions considered in a
continuum media approximation.} magnetization
$\mathbf{M}$ and mobile electron magnetization $\mathbf{m}$, so that the magnetizations
direction can be changed by an external magnetic field $\mathbf{H}$ or
spin-polarized current density $\mathbf{j}$. There is a very thin nonmagnetic
spacer between the layers \textbf{1} and \textbf{2}, which will be considered as a
geometrical plane $x=0$ . To close the electric circuit, a nonmagnetic
metal layer \textbf{3} exists in the region $x\ge L$.

External magnetic field $\mathbf{H}$, lattice and mobile electron
magnetizations of the layer \textbf{1}, $\mathbf{M}_1$ and
$\mathbf{m}_1$, as well as corresponding magnetizations of the
layer \textbf{2}, $\mathbf{M}$ and $\mathbf{m}$ , are assumed
lying in the junction plane $x=0$. Vectors $\mathbf{M}_1$ and
$\mathbf{M}$ make an angle $\chi$ of an arbitrary value under
current $\mathbf{j}=0$. After the current is turned on electrons
transfer from the layer \textbf{1} into the layer \textbf{2}, that
is $j/e>0$. Then the electrons appear in a non-stationary quantum
state in the layer \textbf{2} and "walk" between the spin subbands
of this layer. It corresponds to precession of the mobile electron
magnetization around the lattice magnetization $\mathbf{M}$. We may
introduce two components of the vector $\mathbf{m}$  by an equality
$\mathbf{m}=\mathbf{m}_\|+\mathbf{m}_\perp$, where longitudinal
component $\mathbf{m}_\|$ is parallel to $\mathbf{M}$ and
transversal component $\mathbf{m}_\perp$ is perpendicular to
$\mathbf{M}$. As it was first shown in \cite{Slonczewski,Berger}, angle of
the precession decreases with coordinate $x$ increasing and tends
to zero at $x\approx\lambda_F$, where $\lambda_F\sim 1$ nm is an
electron quantum wave length at the Fermi surface. This may be
valid only because of electron velocity statistical spread and no
relaxation processes are needed to provide such a behavior (more
details may be found in the recent preprint \cite{Gulyaev4}
 also). Therefore, both components $\mathbf{m}_\|$ and $\mathbf{m}_\perp$ exist in the
region $0\le x\le\lambda_F$, but only component  $\mathbf{m}_\|$
remains in the region $x>\lambda_F$ ($\mathbf{m}_\bot=0$).

The region $0\le x\le\lambda_F$ was introduced firstly by
Slonczewski \cite{Slonczewski} and Berger \cite{Berger} and further
we will refer to it as "SB layer". We follow the original works \cite{Slonczewski,Berger} and assume ballistic regime of motion into the SB layer. The
validity criteria for this
assumption may be written as $\lambda_F<l_p$, where momentum free
pass length $l_p$ may be estimated as
$l_p\sim$1--10 nm for metals at room temperature.

\section{Equations}\label{sec3}

We describe the motion of vector $\mathbf{M}$ by means of the Landau--Lifshitz--Gilbert (LLG) equation \cite{Akhiezer}:
\begin{equation}\label{1}
\frac{\partial\mathbf{M}}{\partial
t}+\gamma[\mathbf{M},\mathbf{H}_{eff}]-\frac{\kappa}{M}\Bigl[\mathbf{M},\frac{\partial\mathbf{M}}{\partial
t}\Bigr]=0,
\end{equation}
where $\gamma$ is the gyromagnetic ratio, $\kappa$ is dimensionless damping constant ($0<\kappa\ll 1$),
\begin{equation}\label{2}
  \mathbf{H}_{eff}=\mathbf{H}+\mathbf{H}_a+A\frac{\partial^2\mathbf{M}}{\partial
  x^2}+\mathbf{H}_d+\mathbf{H}_{sd}
\end{equation}
is effective field, $\mathbf{H}_a=\beta\mathbf{M}$ is anisotropy field, $A$ is the intralattice
inhomogeneous exchange constant, $\mathbf{H}_d$ is demagnetization
field, $\mathbf{H}_{sd}$
is \textit{sd} exchange effective field. The latter takes the form~\cite{Akhiezer}
\begin{equation}\label{3}
  \mathbf{H}_{sd}(x,t)=-\frac{\delta U_{sd}}{\delta\mathbf{M}(x,t)},
\end{equation}
where $\delta(\ldots)/\delta\mathbf{M}(x,t)$ is a variational derivative, and $U_{sd}$ is sd exchange energy,
\begin{equation}\label{4}
  U_{sd}=-\alpha\int_0^L\mathbf{m}(x',t)\mathbf{M}(x',t)\,dx',
\end{equation}
parameter $\alpha$ being a dimensionless \textit{sd} exchange constant (typical
estimation is $\alpha\sim 10^4$--10$^6$ ~\cite{Gulyaev1}). Due to the last term in~(\ref{2}), the
motions of $\mathbf{M}$ and $\mathbf{m}$ vectors appear to be coupled.

We describe the motion of vector $\mathbf{m}$  by means of continuity equation (e.g., see~\cite{Aronov,Dyakonov}):
\begin{equation}\label{5}
  \frac{\partial\mathbf{m}}{\partial
  t}+\frac{\partial\mathbf{J}}{\partial
  x}+\gamma\alpha[\mathbf{m},\mathbf{M}]+\frac{\Delta\mathbf{m}}{\tau}=0,
\end{equation}
where $\tau$ is a time of relaxation to the local equilibrium
value $\bar\mathbf{m}\equiv\bar{m}\mathbf{\hat M}$, $\Delta\mathbf{m}\equiv\mathbf{m}-\mathbf{\bar m}=\Delta m\mathbf{\hat M}$ and
$\hat\mathbf{M}\equiv\mathbf{M}/M$ is the unit vector,
$\mathbf{J}$ is mobile electron magnetization flux density.

\section{Vector boundary conditions}\label{sec4}

The SB layer may influence as a special boundary
condition if the thickness of the layer \textbf{2} is large enough, namely, $L\gg\lambda_F$. Moreover, the value
of $\lambda_F$ should be assumed the smallest one among the other lengths in the
system (e.g., $\sqrt A\gg\lambda_F$ and $l\gg\lambda_F$). We may sum the Eqs. (\ref{1}) and (\ref{5}) and so
consider the continuity equation for the total magnetization vector
$\mathbf{M}+\mathbf{m}$. Let us integrate the sum equation over the region $0\le x\le\lambda_F$, the
integration being denoted as
\begin{equation}\label{6}
   \int_{-\varepsilon}^{\lambda_F+\varepsilon}(\ldots)\,dx\equiv\langle\ldots\rangle.
\end{equation}

Then the following three types of summands appear. The first type summands are proportional to small length $\lambda_F$ and may be neglected
for actual values of parameters (fields: $\mathbf{H}$, $\mathbf{H}_a$, $\mathbf{H}_d$; relaxation
parameters: $\kappa$, $\tau$; and frequencies $\omega$ in a microwave region or
lower). The second type of summands may be presented as derivatives
with respect to coordinate $x$. These are spin fluxes of mobile
electron and lattice spins. After integration in (\ref{6}), they do not
depend on $\lambda_F$ and remain finite in the limit $\lambda_F\rightarrow 0$. Finally, the third summand
appears, which is proportional to a singular $\delta$-function inside the SB
layer. This summand arises due to nonequilibrium longitudinal spin
injection by current. After integration in (\ref{6}), it does not depend
on $\lambda_F$ also.

Going the way indicated, we obtain the following vector boundary
condition:
\begin{equation}\label{7}
\mathbf{J}(\lambda_F+\varepsilon)-\mathbf{J}(-\varepsilon)+\mathbf{J}_M(\lambda_F+\varepsilon)-\mathbf{J}_M(-\varepsilon)+\langle\mathbf{G}_{sd}\rangle
=0.
\end{equation}

The mobile electron spin flux $\mathbf{J}$ in the condition (\ref{7}) originates from
the equation (\ref{5}). The lattice magnetization flux
$\mathbf{J}_M$ follows from the intralattice exchange effective
field of Eq. (\ref{1}). The following expression being valid:
\begin{equation}\label{8}
  \mathbf{J}_M=a\bigl[\mathbf{\hat M},\frac{\partial\mathbf{M}}
{\partial x}\bigr]
\end{equation}
and $a=\gamma AM$ is a lattice magnetization diffusion constant.

To calculate the last term $\langle\mathbf{G}_{sd}\rangle$ in
(\ref{7}), we take the sum equation for $\mathbf{M}+\mathbf{m}$ and
get
\begin{equation}\label{9}
\langle\mathbf{G}_{sd}\rangle\equiv\gamma\langle[\mathbf{M},\mathbf{H}_{sd}]\rangle+\gamma\alpha\langle[
\mathbf{m},\mathbf{M}]\rangle=\gamma\alpha\langle\Bigl[
\mathbf{M}(x),\int_0^L(\delta\Delta m/\delta\mathbf{\hat
M})\,dx'\Bigr]\rangle,
\end{equation}
where formulae (\ref{3}) and (\ref{4}) are used under the
calculations. The condition (\ref{7}) remains valid at $x=L$, if we
replace: $\lambda_F+\varepsilon\rightarrow L+\varepsilon$,
$-\varepsilon\rightarrow L-\varepsilon$ and put
$\langle\mathbf{G}_{sd}\rangle\rightarrow 0$.

\section{Nonequilibrium mobile electron spin flux and
magnetization}\label{sec5}

To make the system of equations (\ref{1})--(\ref{5}) and boundary
condition (\ref{7}) be well defined, we should express all the
quantities involved via the vectors $\mathbf{M}$ and $\mathbf{m}$.
As it may be seen from (\ref{7}), we should consider the spin flux $\mathbf{J}$ outside
the SB layer only. In the region $x>\lambda_F$ electrons
occupy subbands having spins parallel to $\mathbf{M}$ ($\uparrow$) and
antiparallel to $\mathbf{M}$ ($\downarrow$). Therefore, the following representations are
valid: $\mathbf{m}=\mu_B(n_\uparrow-n_\downarrow)\mathbf{\hat M}\equiv
m\mathbf{\hat M}$ and $\mathbf{J}=(\mu_B/e)(j_\uparrow-j_\downarrow)\mathbf{\hat
M}$, where $\mu_B$ is Bohr magneton,
$n_{\uparrow,\downarrow}$ and $j_{\uparrow,\downarrow}$ are partial electron densities and current densities in the
spin subbands, respectively. The total electron density $n=n_\uparrow
+n_\downarrow$ and $j=j_\uparrow+j_\downarrow$ do not depend on $x$ and $t$
because of local
neutrality conditions in metal and one-dimensional geometry of our model. We use the well-known "diffusion-drift" formula for $j_{\uparrow,\downarrow}$  to express
the partial currents via $j$, $n$  and $\mathbf{m}$. The corresponding calculations
are direct and were done completely in \cite{Gulyaev1}. The final result
for $x>\lambda_F$ may be written as follows
\begin{equation}\label{10}
\mathbf{J}=\Bigl(\frac{\mu_B}{e}Qj-\tilde D\frac{\partial m}{\partial x}\Bigr)\mathbf{\hat
M},
\end{equation}
where $Q=(\sigma_\uparrow-\sigma_\downarrow)/(\sigma_\uparrow+\sigma_\downarrow)$
may be understood as an equilibrium conductivity spin
polarization parameter with $\sigma_{\uparrow,\downarrow}\equiv\mu_{\uparrow,\downarrow}n_{\uparrow,\downarrow}$, and  
$\tilde D=(\sigma_\uparrow D_\downarrow+\sigma_\downarrow D_\uparrow)/
(\sigma_\uparrow+\sigma_\downarrow)$
is the effective spin diffusion constant, quantities $\mu_{\uparrow,\downarrow}$ and $D_{\uparrow,\downarrow}$ being partial mobilities and diffusion constant, respectively. To obtain (\ref{10}), an additional assumption should
be made \cite{Gulyaev1}, namely, $j/j_D\ll 1$, where $\j_D\equiv enl/\tau$ is a characteristic current density in
the layer \textbf{2}. With typical parameter values, $n\sim
10^{22}$ cm$^{-3}$, $l\sim 3\times 10^{-6}$ cm, $\tau\sim 3\times
10^{-13}$ s, we get $j_D\sim 1.6\times 10^{10}$ A/cm$^2$.

The lattice in the layer \textbf{1} is pinned. Therefore, the magnetization
flux $\mathbf{J}(-\varepsilon)$ may be written similar to
(\ref{10})
but without the spatial derivative. We introduce $\mathbf{\hat
M}_1=\mathbf{M}_1/M_1$ and have
\begin{equation}\label{11}
{\mathbf{J}}\left( { - \varepsilon } \right) = \frac{{\mu _B }}
{e}Q_1 j{\mathbf{\hat M}}_1.
\end{equation}

The latter flux has longitudinal and transversal components, which are:
\begin{eqnarray}\label{12}
  &&J_{||} \left( { - \varepsilon } \right) = \frac{{\mu _B }}
{e}Q_1 j\left( {{\mathbf{\hat M}}_1 {\mathbf{\hat M}}\left( \varepsilon  \right)} \right){\mathbf{\hat M}}\left( \varepsilon
\right)\nonumber \\ &&\mathrm{and} \\ &&J_ \bot  \left( { - \varepsilon } \right) = \frac{{\mu _B }}
{e}Q_1 j\left[ {{\mathbf{\hat M}}\left( \varepsilon  \right),\left[ {{\mathbf{\hat M}}_1 ,{\mathbf{\hat M}}\left( \varepsilon  \right)} \right]}
\right].\nonumber
\end{eqnarray}

We should calculate now the nonequilibrium magnetization for $x>\lambda_F$. In
the region an effective frequency of the motion $\omega$ is determined by
the precession in relatively small fields: $\omega\sim\gamma H$, $\gamma H_d$, $\gamma H_a\ll\gamma\alpha M\equiv\omega_{sd}$. We assume the
conditions $\omega\tau\ll 1$ and $\omega_{sd}\tau\gg 1$ are valid for typical values $\tau\sim 3\times 10^{-13}$ s, $\alpha\sim2\times 10^4$, $M\sim 10^3$ G. It
allows neglect the time derivative in (\ref{5}) and substitute the
formula (\ref{10}) for flux $\mathbf{J}$. Then the equation reduces to
\begin{equation}\label{13}
  \frac{\partial^2m}
{\partial x^2}-\frac{\Delta m}
{l^2} = 0,
\end{equation}
where $l=\sqrt{\tilde D\tau}$ is the spin diffusion length.
Analogous relations may be written, of course, for the nonmagnetic layer
\textbf{3} with the following modifications: $\bar m_3=0$ and $Q_3=0$.

The next step is to find the solution of (13) that satisfies the conditions:
1) $J_{||} \left( { - \varepsilon } \right) = J\left( {\lambda _F  + \varepsilon } \right)
$, 2) $J\left( {L - \varepsilon } \right) = J\left( {L + \varepsilon } \right)
$, 3) continuity of subband chemical potential difference at the interface point $x=L$  (see \cite{Gulyaev4,Gulyaev5} for details). This solution takes the form
\begin{eqnarray}\label{14}
&&\Delta m(x)=\frac{j}{j_D}\frac{\mu_Bn}
{\sinh\lambda+\nu\cosh\lambda}\{Q\cosh\xi+[Q_1(\mathbf{\hat
M}_1\mathbf{\hat M}(\varepsilon))-Q]\nonumber \\
&&\times[\cosh(\lambda-\xi)+\nu\sinh(\lambda- \xi)]\},
\end{eqnarray}
where $\lambda=L/l$, $\xi=x/l$ and parameter $\nu$ characterizes the
influence of the layer \textbf{3} (typically $\nu\sim 1$, see
\cite{Gulyaev4,Gulyaev5} for details). We substitute (\ref{14})
into expression (\ref{9}) and use the variational derivative $\delta(\mathbf{\hat M}_1\mathbf{M}(\varepsilon))/\delta\mathbf{M}(x)=\mathbf{\hat M}_1\delta(x-\varepsilon)$.
Then we get
\begin{equation}\label{15}
\langle\mathbf{G}_{sd}\rangle=\gamma\alpha\mu_BnQ_1l\frac{j}{j_D}\left(1-\frac{\nu}{\sinh\lambda+\nu\cosh\lambda}\right)\left[\mathbf{M}(\varepsilon),\hat{\mathbf{M}}_1\right],
\end{equation}
because of $\langle\delta(x-\varepsilon)\rangle=1$.

Turning back to condition (\ref{7}), we see all the terms are now
explicitly dependent on the vectors $\mathbf{M}$ and $\mathbf{m}$. The latter vectors
satisfy the equation of motion (\ref{1})--(\ref{5}). Therefore, the way opens
now for solving our problem directly.

\section{Initial steady state and fluctuation
instability}\label{sec6}

To illustrate the possibilities of solution, we assume simple
situation where field $\mathbf{H}$ is applied along the positive direction of
$z$ axis, $\mathbf{H}_a$ is parallel to this axis too and the angle $\chi=\pi$. In the
situation all the equations and boundary conditions are satisfied
for the following steady state magnetization vector $\mathbf{\hat
M}$: $\hat{\bar M}_x=\hat{\bar M}_y=0$, $\hat{\bar
M}_z=1$, and $\mathbf{\hat M}_1=-\mathbf{\hat z}$.

We introduce the fluctuations $\Delta\mathbf{\hat M}$ by the equality: $\mathbf{\hat M}=\mathbf{\hat z}+\Delta\mathbf{\hat M}$ with $\mathbf{\hat M}\Delta\mathbf{\hat M}=0$ and $\Delta\hat M\equiv|\Delta\mathbf{\hat M}|\ll 1$. We
linearize the equations LLG and condition (\ref{7}) with respect to $\Delta\mathbf{\hat M}$  and
lay $\mathbf{H}_d=-4\pi M\Delta\hat M_x\mathbf{\hat x}$. Then we get the equations for
fluctuations $\Delta\hat M_{x,y}\sim\exp(iqx-i\omega t)$:
\begin{eqnarray}\label{16}
  \frac{\partial^2\Delta\hat M_x}
{\partial x^2}-\frac{\left(\Omega_x-i\kappa\omega\right)}
{a}\Delta\hat M_x-\frac{i\omega}
{a}\Delta\hat M_y= 0,\nonumber \\  \\
\frac{\partial^2\Delta\hat M_y}
{\partial x^2}-\frac{\left(\Omega_y-i\kappa\omega\right)}
{a}\Delta\hat M_y+\frac{i\omega}
{a}\Delta\hat M_x= 0,\nonumber
\end{eqnarray}
where two characteristic frequencies appear: $\Omega _x  = \gamma \left( {H + H_a  + 4\pi M} \right)$, $\Omega _y  = \gamma \left( {H + H_a } \right)$ and transversal
part of the vector boundary condition (\ref{7}) for fluctuations at
$x=0$
takes the form:
\begin{eqnarray}\label{17}
  \frac{{\partial \Delta \hat M_x }}
{{\partial x}} = k\Delta \hat M_y  - p\Delta \hat M_x ,\nonumber \\
\\ \frac{{\partial \Delta \hat M_y }}
{{\partial x}} =  - k\Delta \hat M_x  - p\Delta \hat M_y .\nonumber
\end{eqnarray}
Parameters $k = (\mu _B jQ_1 )/(e\gamma M^2 A)$
and $p=\alpha\mu_BjQ_1\tau[1-\nu/(\sinh\lambda+\nu\cosh\lambda)]/
(eAM)$ are proportional to current density $j$ and
describe respectively the effects of spin-transfer torque and spin
injection.

Transversal terms of the boundary condition (\ref{7}) at $x=L$ become:
\begin{equation}\label{18}
  \frac{\partial\Delta\hat M_x}{\partial x}=0,\quad \frac{\partial\Delta\hat M_y}{\partial
  x}=0.
\end{equation}

Solving the standard boundary problem (\ref{16})--(\ref{18}), we
come to the characteristic wave number
\begin{equation}\label{19}
  q^2=\frac{1}{2a}\bigl[\Omega_x+\Omega_y-2i\kappa\omega
\pm\sqrt{(\Omega_x-\Omega_y)^2+4\omega^2}\bigr]
\end{equation}
and dispersion relation
\begin{equation}\label{20}
 qL\tan qL =  - pL \pm ikL
\end{equation}
for determining the spin-wave spectrum and decrement $\omega(q)$.

\section{Steady state arising due to instability
development}\label{sec7}

The dispersion relation (\ref{20}) is the same as in our previous
works \cite{Gulyaev4,Gulyaev5} and therefore all the results about the instability
thresholds and current dependent spectrum remain valid. But in our
new approach, contrary to \cite{Gulyaev4,Gulyaev5}, we have much more simple
equations of motion, which do not contain any singular terms. It
allows solve more complicated problems now. One of such problem is:
what will be a steady state the instability development leads to?

We confine ourselves to the case of large enough thickness $L>l$,
$\sqrt A$ corresponding to the injection mechanism dominated. In such case,
the only polar angle $\theta (x)$ exists that determines the stationary
magnetization $\mathbf{\hat M}$  orientation and the problem reduces to solving of
nonlinear equation
\begin{equation}\label{21}
  \delta ^2 \frac{{d^2 \theta }}
{{dx^2 }} - \sin \theta \cos \theta  - \frac{H}
{{H_a }}\sin \theta  = 0,
\end{equation}
where $\delta=\sqrt{A/\beta}$. Transversal vector boundary conditions (\ref{7}) in the polar
coordinates are
\begin{equation}\label{22}
  \frac{d\theta}{dx}+p\sin\theta\Bigr|_{x=0}=0\quad
  \mathrm{and}\quad\frac{d\theta}{dx}\Bigr|_{x=L}=0.
\end{equation}

Equation (\ref{21}) describes a "classical" particle moving in a
potential field. As it is well known, such type of equation may be
integrated easily. At $p\delta<\left(1+H/H_a\right)^{1/2}$, there is no instability and only one
solution $\theta=0$ exists. Under the opposite condition, the
instability occurs and a nonzero solution appears which describes
the new steady state orientation angle $\theta(x)$ in the layer \textbf{2}:
\begin{eqnarray}\label{23}
&\theta(x)=2\arccos\Biggl\{\Bigl((hj/j_0)\tanh(x/h\delta)+1\Bigr)(1-h^2)^{1/2}
  \nonumber \\
  &\times\Bigl[\bigl(hj/j_0+\tanh(x/h\delta)\bigr)^2-h^2\bigl((hj/j_0)\tanh(x/h\delta)+1\bigr)^2\Bigr]^{-1/2}\Biggr\},
\end{eqnarray}
where $h=\left(1+H/H_a\right)^{-1/2}$,\quad $j_0=e\delta
H_a/\alpha\mu_B\tau Q_1$.

\section*{Acknowledgements}

The work was supported by RFBR grants Nos. 03-02-17540 and 04-02-08248.

\end{document}